\documentclass[12pt]{article}
\usepackage{epsf}
\usepackage{amsmath, amssymb}

\begin{document}

\renewcommand{\theequation}{\thesection.\arabic{equation}}

\renewcommand{\thefootnote}{\fnsymbol{footnote}}
\setcounter{footnote}{0}

\begin{titlepage}

\def\thefootnote{\fnsymbol{footnote}}

\begin{center}

\hfill UT-12-23\\
\hfill July, 2012\\

\vskip .75in

{\Large \bf 

Thermal Effects on Saxion in Supersymmetric Model
\\
with Peccei-Quinn Symmetry

}

\vskip .75in

{\large
Takeo Moroi and Masahiro Takimoto
}

\vskip 0.25in

{\em Department of Physics, University of Tokyo,
Tokyo 113-0033, Japan}

\end{center}
\vskip .5in

\begin{abstract}

  We consider supersymmetric model with Peccei-Quinn symmetry and
  study effects of saxion on the evolution of the universe, paying
  particular attention to the effects of thermal bath.  The axion
  multiplet inevitably couples to colored particles, which induces
  various thermal effects.  In particular, (i) saxion potential is
  deformed by thermal effects, and (ii) coherent oscillation of the
  saxion dissipates via the interaction with hot plasma.  These may
  significantly affect the evolution of the saxion in the early
  universe.

\end{abstract}

\end{titlepage}

\renewcommand{\thepage}{\arabic{page}}
\setcounter{page}{1}
\renewcommand{\thefootnote}{\#\arabic{footnote}}
\setcounter{footnote}{0}

\section{Introduction}
\label{sec:introduction}
\setcounter{equation}{0}

Peccei-Quinn (PQ) mechanism \cite{Peccei:1977hh, Peccei:1977ur}
provides an elegant solution to the strong CP problem.  Promoting the
$\theta$-parameter to a dynamical variable, $\theta=0$ is realized
after the spontaneous breaking of the PQ symmetry.  In addition, in
such a framework, a very light scalar field, called axion, shows up
\cite{Weinberg:1977ma, Wilczek:1977pj}.  The coherent oscillation of
the axion field is a viable candidate of dark matter of our universe
if the PQ symmetry breaking scale is around $10^{12}\ {\rm GeV}$
\cite{Nakamura:2010zzi}.  Thus, the PQ mechanism has been attracted
many attentions not only from particle-physics point of view but also
from cosmology point of view.

If we embed the PQ mechanism into the framework of supersymmetry
(SUSY), which is a prominent candidate of the model beyond the
standard model, serious cosmological difficulties may show up.  In
particular, because of the supersymmetry, there should exist
superpartners of the axion, i.e., axino and saxion.  Thermally
produced axino and saxion, as well as coherent oscillation of saxion,
may significantly affect the evolution of the universe.  In order to
construct cosmologically viable supersymmetric PQ model, it is
important to understand how the axion multiplet affects the thermal
history of the universe.

The purpose of this paper is to study the effect of the saxion on
cosmology, paying particular attention to the thermal effects on
properties of the saxion.  In hadronic axion models \cite{Kim:1979if,
  Shifman:1979if}, axion multiplet couples to extra colored
multiplets, which include colored fermions (called PQ fermions), to
solve the strong CP problem.  Even in DFSZ-type model
\cite{Dine:1981rt, Zhitnitsky:1980tq} in which the ordinary Higgses
have non-vanishing PQ charges, extra colored multiplets are likely to
be introduced to avoid domain-wall problem. The saxion acquires
thermal mass of the order of $yT$ (with $y$ being coupling constant)
when the cosmic temperature $T$ is higher than the mass of PQ
fermions.  Even if the temperature is lower so that the densities of
the PQ fermions (and their scalar partners) are negligibly small,
thermal effect induces logarithmic term in saxion potential
\cite{Anisimov:2000wx}.  These thermal effects shift the minimum of
the potential, which may result in too large amplitude of the coherent
oscillation of the saxion.  Furthermore, the coherent oscillation of
the saxion may dissipate via the interaction with hot QCD plasma.
Taking these effects into account, we study cosmological history of a
SUSY PQ model.

\section{Thermal Effects on Saxion}
\label{sec:thermal}
\setcounter{equation}{0}

In supersymmetric PQ model, there exists supermultiplet $\hat{\cal A}$
(called ``axion supermultiplet'') which includes the axion field.
Saxion (denoted as $\sigma$) is a real scalar field in the axion
multiplet:
\begin{eqnarray}
  \hat{\cal A} = \frac{1}{\sqrt{2}} (\sigma + i a)
  + \sqrt{2} \theta \tilde{a} + (F\mbox{-term}),
\end{eqnarray}
where $a$ and $\tilde{a}$ are axion and axino, respectively.  (Here
and hereafter, the ``hat'' is for superfields.)  The saxion potential
is lifted only by the supersymmery breaking effects, and hence is very
flat at $T=0$.

The axion multiplet couples to colored multiplets, which include PQ
fermions, to solve the strong CP problem.  In this study, we consider
the case that the masses of PQ fermions are related to the amplitude
of $\sigma$.  The PQ fermions become massless when $\sigma$ takes a
particular value.  By integrating out PQ fermions, the axion multiplet
couples to gauge multiplets (in particular, that of $SU(3)_C$) as
\begin{eqnarray}
  {\cal L}_{\rm int} = 
  \frac{\alpha_3}{8\pi F_a}
  \int d^2 \theta \hat{\cal A} \hat{\cal W}^\alpha \hat{\cal W}_\alpha 
  + {\rm h.c.},
  \label{AWW}
\end{eqnarray}
where $\hat{\cal W}$ is the field strength supermultiplet of
$SU(3)_C$.  (Here and hereafter, $\alpha_i=g_i^2/4\pi$, with $g_i$
being the gauge coupling constant of $U(1)_Y$, $SU(2)_L$, $SU(3)_C$
for $i=1$, $2$, and $3$, respectively)

The potential of the saxion is affected by particles in thermal bath
(in particular, by those with $SU(3)_C$ quantum numbers).  In
addition, scattering processes may also become important.  In the
following, we discuss possible thermal effects on the saxion.

\subsection{Thermal Potential}

The first subject is thermal effects on the potential (more precisely,
the free energy).  The PQ fermions (and their superpartners), which we
call $Q$, affect the potential as
\begin{eqnarray}
  V_T (\sigma) = V_0 (\sigma)
  + g_Q^{\rm (B)} V_T^{\rm (B)} (m_Q(\sigma)) 
  + g_Q^{\rm (F)} V_T^{\rm (F)} (m_Q(\sigma)),
  \label{V_T(general)}
\end{eqnarray}
where $V_0$ is the zero-temperature potential while $V_T^{\rm (B)}$
and $V_T^{\rm (F)}$ are contributions of bosonic and fermionic fields,
respectively, and $m_Q(\sigma)$ is the mass of the fields which couple
to $\sigma$.  (Extra contributions exist if particles other than PQ
fermions couple to $\sigma$.)  At the one-loop level, the functions $
V_T^{\rm (B/F)}$ are given by \cite{Dolan:1973qd}
\begin{eqnarray}
  V_T^{\rm (B/F)} (M) = \pm
  \frac{T^4}{\pi^2} \int_0^\infty dz 
  z^2 \ln \left[ 1 \mp e^{-\sqrt{z^2 + M^2/T^2}} \right],
\end{eqnarray}
where upper and lower signs are for $V_T^{\rm (B)}$ and $V_T^{\rm
  (F)}$, respectively.  In addition, $g_Q^{\rm (B/F)}$ are the number
of degrees of freedom.  ($g_Q^{\rm (B)}=1$ for one complex scalar
field, and $g_Q^{\rm (F)}=1$ for one chiral fermion.)

$V_T^{\rm (B)}$ and $V_T^{\rm (F)}$ change their behavior at $M\sim
T$.  When $M\lesssim T$, $V_T^{\rm (B)}$ and $V_T^{\rm (F)}$ are well
approximated as
\begin{eqnarray}
  V_T^{\rm (B)} (M) &\simeq& 
  -\frac{\pi^2}{45} T^4 + \frac{1}{12} T^2 M^2 + O(TM^3),
  \\
  V_T^{\rm (F)} (M) &\simeq& 
  -\frac{7\pi^2}{360} T^4 + \frac{1}{24} T^2 M^2 + O(M^4),
\end{eqnarray}
and hence the so-called thermal mass shows up.  When $\sigma$
satisfies the condition $m_Q(\sigma)\lesssim T$, $V_T(\sigma)$
acquires a new term of $\sim T^2 m_Q^2(\sigma)$.  For $M\gtrsim T$, on
the contrary, $V_T^{\rm (B)}$ and $V_T^{\rm (F)}$ rapidly go to zero.

If the temperature is high enough so that $T\gg m_Q(\sigma_0)$, where
$\sigma_0$ is the zero-temperature vacuum expectation value (VEV) of
$\sigma$, the PQ fermions are in thermal bath as relativistic
particles even when $\sigma\sim\sigma_0$.  In such a case, the
expectation value of $\sigma$ is determined by the thermal mass term.
Because $V_T^{\rm (B)} (m_Q(\sigma))$ and $V_T^{\rm (F)}
(m_Q(\sigma))$ are minimized at $\sigma=\sigma_Q$ (where $\sigma_Q$
satisfies $m_Q(\sigma_Q)=0$), we expect only one minimum of the
potential at $\sigma\sim\sigma_Q$ at high enough temperature.  The
situation changes when the temperature is lower than $m_Q(\sigma_0)$.
In such a case, the number densities of $Q$ are Boltzmann suppressed
at $\sigma\sim\sigma_0$.  Thermal effects are negligible around the
zero-temperature minimum and there exist a minimum at
$\sigma\sim\sigma_0$.  (Effect of the logarithmic thermal correction
to the potential \cite{Anisimov:2000wx}, which affects the position of
the minimum, may be sizable.  See the discussion below.)  On the other
hand, if $V_0(\sigma)$ is flat enough, the minimum at
$\sigma\sim\sigma_Q$ also remains.  In such a case, there are two
distinctive minima of the potential; even though there exists a
minimum at $\sigma\sim\sigma_0$, the scalar field $\sigma$ can be
thermally trapped in the different minimum close to the symmetry
enhanced point.  Then, the minimum at $\sigma\sim\sigma_Q$ disappears
when the temperature is so low that the shape of the potential around
$\sigma\sim\sigma_Q$ is determined by the zero-temperature potential.

The saxion potential has another important term which is not taken
into account in Eq.\ \eqref{V_T(general)}.  Because the free energy of
hot QCD plasma has a contribution proportional to $g_3^2(T) T^4$
\cite{Kajantie:2002wa}, and also because $g_3^2(T)$ depends on
$m_Q(\sigma)$ (i.e., the mass of colored particles) if
$m_Q(\sigma)>T$, the free energy has the following term
\cite{Anisimov:2000wx}\footnote
{The logarithmic term is expected to disappear in the region where
  $m_Q(\sigma)$ is smaller than $\sim T$.  In such a region, however,
  $V_T^{\rm (B)}$ and $V_T^{\rm (F)}$ dominate, so the effect of the
  logarithmic term is not important.}
\begin{eqnarray}
  V_L (\sigma) \equiv a_L \alpha_3^2 (T) T^4 \ln |m_Q(\sigma)|^2,
\end{eqnarray}
where $a_L$ is a constant a bit larger than $1$.

\subsection{Scattering Processes and Dissipation}

Next, we consider the scattering processes and dissipation.

It has been known that scatterings of thermal particles contribute to
the production processes of axion, axino and saxion.  Define the yield
variable as
\begin{eqnarray}
  Y_X \equiv \frac{n_X}{s},
\end{eqnarray}
where $n_X$ is the number density of particle $X$ while
$s=\frac{45}{2\pi^2} g_*(T) T^3$ is the entropy density (with $g_*$
being the effective number of massless degrees of freedom).  Then, if
the interaction of the axion multiplet with the minimal supersymmetric
standard model (MSSM) particles is dominated by the operator given in
Eq.\ \eqref{AWW}, thermally produced axino abundance in
radiation-dominated era is given by \cite{Brandenburg:2004du}
\begin{eqnarray}
  \left[ Y_{\tilde{a}}^{\rm (th)} \right]_{\rm RD}
  \simeq
  \mbox{min} \left[
    Y_{\tilde{a}}^{\rm (eq)},
    0.20 \times \alpha_3^3 
    \ln \left( \frac{0.0977}{\alpha_3} \right)
    \left( \frac{T_{\rm R}}{10^{7}\ {\rm GeV}} \right)
    \left( \frac{F_a}{10^{11}\ {\rm GeV}} \right)^{-2}
  \right],
  \label{Y(axino)}
\end{eqnarray}
where $T_{\rm R}$ is the reheating temperature, and
$Y_{\tilde{a}}^{\rm (eq)}\simeq 1.8\times 10^{-3}$ is the thermal
abundance of axino.  A precise calculation of the abundance of
thermally produced saxion is not available yet.  Here, using the fact
that the production processes of saxion and axino are governed by the
same supersymmetric interaction, we approximate the thermally produced
saxion abundance as $[ Y_{\sigma}^{\rm (th)} ]_{\rm RD}\simeq
\frac{2}{3}[ Y_{\tilde{a}}^{\rm (th)}]_{\rm RD}$.

The thermally produced axino may be cosmologically harmful
\cite{Covi:2001nw}.  If it is the lightest superparticle (LSP) and is
stable, it contributes to the present dark matter density.  In such a
case, a very stringent upper bound on the reheating temperature is
obtained in order not to overproduce axino.  (For recent discussion,
see, for example, \cite{Bae:2011jb, Choi:2011yf} and references
therein.)  Even if the saxion is unstable, the LSP produced by the
decay of axino contributes to the present mass density of the
universe.  If all the LSPs produced by the axino decay survive until
today, low reheating temperature is needed to avoid the overproduction
of the LSP.  In the case of unstable axino, these cosmological
difficulties can be avoided if the LSP has large enough pair
annihilation cross section \cite{Choi:2008zq}.  This may be the case
if, for example, the LSP is neutral Wino \cite{Moroi:1999zb,
  Baer:2011hx, Baer:2011uz}.  In addition, $R$-parity violation may be
another possibility to avoid the cosmological difficulty.

Existence of hot plasma not only increases but also reduces the number
density of particles in the PQ sector.  Such an effect is particularly
important in studying the evolution of saxion oscillation.  The
dissipation rate of the saxion is related to the bulk viscosity of hot
plasma \cite{Bodeker:2006ij, Laine:2010cq}, and is estimated
as\footnote
{The authors thank K. Mukaida for helpful discussion on this issue.}
\begin{eqnarray}
  \Gamma_{\rm diss} \sim \frac{9\alpha_3^2}{128\pi^2 \ln\alpha_3^{-1}}
  \frac{T^3}{F_a^2},
  \label{Gamma_diss}
\end{eqnarray}
where, if the saxion is far away from the minimum, $F_a$ should be
replaced by the effective axion decay constant $F_a^{\rm (eff)}$
(which is $\sim \sigma$).  Notice that, if $\sigma$ interacts with
particles other than MSSM gauge multiplets, $\Gamma_{\rm diss}$
receives contributions from those extra particles.  We will see that
this happens in some case.

The dissipation rate $\Gamma_{\rm diss}$ is an important quantity in
studying the saxion oscillation in the early universe because
$\Gamma_{\rm diss}$ may become larger than the expansion rate of the
universe $H$.  For the case that $\Gamma_{\rm diss}\ll H$, the saxion
oscillates if the initial amplitude is non-vanishing.  Then, with
parabolic potential, for example, the energy density of the saxion
decreases as $a^{-3}$ with the cosmic expansion (where $a$ is the
scale factor).  On the contrary, if $\Gamma_{\rm diss}$ is sizable, we
cannot neglect the effect of dissipation.  In particular, if
$\Gamma_{\rm diss}\gtrsim H$, coherent oscillation of the saxion
dissipates within a time scale shorter than the cosmic time.  In
radiation dominated universe with $g_*(T_{\rm R})=228.75$,
$\Gamma_{\rm diss}$ given in Eq.\ \eqref{Gamma_diss} becomes larger
than $H$ when $T\gtrsim 3\times 10^5\ {\rm GeV}$ ($5\times 10^7\ {\rm
  GeV}$, $7\times 10^9\ {\rm GeV}$, $1\times 10^{12}\ {\rm GeV}$) for
$F_a=10^9\ {\rm GeV}$ ($10^{10}\ {\rm GeV}$, $10^{11}\ {\rm GeV}$,
$10^{12}\ {\rm GeV}$).  One should note that, if $\Gamma_{\rm
  diss}\gtrsim H$, $Y_{\sigma}^{\rm (th)}$ becomes comparable to
$Y_{\sigma}^{\rm (eq)}$.  Thus, in such a case, the particles in the
axion multiplet are fully thermalized.

Cosmology of models with saxion was considered in the framework in
which the saxion field plays the role of the flaton field for thermal
inflation \cite{Chun:2000jr, Chun:2000jx, Kim:2008yu}.  We will
consider a different class of models, taking account of the thermal
effects discussed above.  We will see that the evolution of the saxion
in the early universe can be significantly affected by these effects.

\section{Explicit Example}
\label{sec:model}
\setcounter{equation}{0}

\subsection{Model}

To see the thermal effects on the evolution of the saxion, we consider
the supersymmetric PQ model with the following superpotential:
\begin{eqnarray}
  W = W_{\rm MSSM} + \lambda \hat{S} (\hat{\bar{X}} \hat{X} - f^2)
  + y \hat{\bar{X}} \hat{Q}_0 \hat{Q}_+,
  \label{superpot}
\end{eqnarray}
where $W_{\rm MSSM}$ is the superpotential of the MSSM, and $\hat{X}$
($+1$), $\hat{\bar{X}}$ ($-1$), $\hat{Q}_0$ ($0$), $\hat{Q}_+$ ($+1$),
and $\hat{S}$ ($0$) are chiral superfields.  (We denote the $U(1)_{\rm
  PQ}$-charge in the parentheses.)  $\hat{Q}_0$ and $\hat{Q}_+$ are in
the ${\bf 3}$ and ${\bf \bar{3}}$ representations of $SU(3)_C$; here
one pair of PQ fermions are introduced.  One combination of $\hat{X}$
and $\hat{\bar{X}}$ becomes the axion supermultiplet.  In addition,
$y$, $\lambda$, and $f$ are constants which are chosen to be real and
positive.

In the minimum of the supersymmetric scalar potential, we obtain the
relation $\bar{X}X=f^2$.  Such a constraint eliminates one combination
of $\hat{\bar{X}}$ and $\hat{X}$ from the low-energy spectrum and
$\hat{\bar{X}}$ and $\hat{X}$ are decomposed into the axion multiplet
$\hat{\cal A}$ and heavy multiplet which we call $\hat{\cal X}$.
Without SUSY breaking terms, the relative size of $\bar{X}$ and $X$ is
undetermined; such a flat direction corresponds to $\hat{\cal A}$.
The SUSY breaking terms relevant for our study are\footnote
{The tri-linear interaction $S\bar{X}X$ and linear term in $S$ may
  also exist in $V_{\rm soft}$, with which the axino mass is
  generated.  In the present model, the axino mass is at most
  $\frac{1}{\sqrt{2}}m_\sigma$.}
\begin{eqnarray}
  V_{\rm soft} = m_1^2 |X|^2 + m_2^2 |\bar{X}|^2 + m_S^2 |S|^2,
  \label{V_soft}
\end{eqnarray}
where we expect that $m_1$, $m_2$, and $m_S$ are of the order of the
gravitino mass $m_{3/2}$.  With the SUSY breaking terms, the VEVs of
scalar fields are fixed.

At the minimum of the potential, the axion multiplet $\hat{\cal A}$
and the heavy multiplet $\hat{\cal X}$ are embedded into
$\hat{\bar{X}}$ and $\hat{X}$ as
\begin{eqnarray}
  \hat{\bar{X}} &=& F_a \cos\beta_X 
  - \hat{\cal A} \cos\beta_X + \hat{\cal X} \sin\beta_X,
  \\
  \hat{X} &=& F_a \sin\beta_X 
  + \hat{\cal A} \sin\beta_X + \hat{\cal X} \cos\beta_X,
\end{eqnarray}
where $F_a$ is the axion decay constant which is given by
\begin{eqnarray}
  F_a^2 = \langle \bar{X} \rangle^2 + \langle X \rangle^2
  = \frac{2 f^2}{\sin 2\beta_X},
\end{eqnarray}
with
\begin{eqnarray}
  \cos 2\beta_X \equiv \xi \equiv
  \frac{m_1^2 - m_2^2}{m_1^2 + m_2^2}.
  \label{beta_X}
\end{eqnarray}
By integrating out $\hat{Q}_0$ and $\hat{Q}_+$, we obtain the
interaction given in Eq.\ \eqref{AWW}.  Then, the saxion mass is
$m_\sigma = {2m_1 m_2}/{\sqrt{m_1^2 + m_2^2}}$.

We note here that, in the present setup, the decay rate of saxion
$\Gamma_{\sigma}$ is governed by the $\xi$ parameter defined in Eq.\
\eqref{beta_X}.  In hadronic axion model in which axion multiplet does
not couple to the Higgs multiplet, $\sigma$ dominantly decays into
axion pair; decay rate of such a process is given by
\begin{eqnarray}
  \Gamma_{\sigma\rightarrow aa} = 
  \frac{\xi^2}{64\pi} \frac{m_\sigma^3}{F_a^2}.
  \label{Gamma_saxion}
\end{eqnarray}
The saxion also decays into gauge-boson pairs (in particular, the
gluon pair) via the interaction given in Eq.\ \eqref{AWW}.  However,
the saxion-gluon-gluon interaction is loop suppressed and hence such a
process is expected to be subdominant (as far as $\xi\sim 1$); indeed,
the decay rate for the process $\sigma\rightarrow gg$ (with $g$ being
the gluon) is given by
\begin{eqnarray}
  \Gamma_{\sigma\rightarrow gg} = 
  \frac{8 \alpha_3^2}{256\pi^3} \frac{m_\sigma^3}{F_a^2}.
  \label{Gamma_saxion2gluglu}
\end{eqnarray}
Numerically, $\Gamma_{\sigma\rightarrow aa}>\Gamma_{\sigma\rightarrow
  gg}$ when $\xi\gtrsim 0.05$.

$X$ or $\bar{X}$ may directly couple to the MSSM Higgses if they have
non-vanishing PQ charges, which gives the decay mode of saxion into
the Higgs boson pair.  For example, we may introduce the following
term into the superpotential:\footnote
{If $X$, instead of $\bar{X}$, couples to the Higgses, the potential
  at $X\gg f$ may be modified due to extra thermal effect, which may
  change the following discussion.  We do not consider such a case.}
\begin{eqnarray}
  W_{\bar{X}^2H_uH_d} = 
  \frac{\lambda'}{M_{\rm Pl}} \hat{\bar{X}}^2 \hat{H}_u \hat{H}_d,
  \label{XXHH}
\end{eqnarray}
where $\hat{H}_u$ and $\hat{H}_d$ are up- and down-type Higgses,
respectively.  Using the fact that the so-called $\mu$-parameter is
generated by the VEV of $\bar{X}$, we obtain the decay rate as
\begin{eqnarray}
  \Gamma_{\sigma\rightarrow hh} = 
  \frac{1}{2\pi} \frac{m_\sigma^3}{F_a^2}
  \left( \frac{\mu^2}{m_\sigma^2} \right)^4
  \left( 1 - \frac{m_h^2}{m_\sigma^2} \right)^{1/2}.
\end{eqnarray}
$Br(\sigma\rightarrow hh)$ depends on various parameters.  If $\mu\sim
O(m_\sigma)$, $Br(\sigma\rightarrow hh)$ and $Br(\sigma\rightarrow
aa)$ are of the same order.  In models with smaller $m_{\sigma}$ (like
the gauge mediation), the decay process of $\sigma\rightarrow hh$ is
kinematically blocked.

Since we are interested in the case of $F_a\sim 10^{9-12}\ {\rm GeV}$,
the lifetime of the saxion becomes very long.  If there exists relic
saxion in the early universe, the decay of saxion produces axion which
survives until today and behaves as an extra radiation component
(so-called ``dark radiation'') \cite{Chun:2000jr}.  We will study
implications of such a relativistic axion in the present model.

\subsection{Thermal Potential}

Now, we are at the position to discuss the potential of the saxion.
We first comment on the so-called Hubble-induced mass.  In the
inflaton dominated era before reheating, Planck suppressed
interactions of $\hat{\bar{X}}$ and $\hat{X}$ may induce their
effective masses of the order of the expansion rate of the universe.
Even in radiation dominated universe, this may be the case
\cite{Kawasaki:2011zi}.  (We call such masses as Hubble-induced
masses.)  The sizes of the Hubble-induced masses crucially depend on
the model, and we simply parametrize the potential as\footnote
{If $\bar{X}$ or $X$ couples to extra fields, contributions of those
  extra fields should be taken into account.  This is the case if, for
  example, $\hat{Q}_0$ and $\hat{Q}_+$ are embedded into complete
  multiplets of the gauge group of grand unification.}
\begin{eqnarray}
  V_T &=& \tilde{m}_1^2 |X|^2 + \frac{\tilde{m}_2^2 f^4}{|X|^2}
  + 2 N_C V_T^{\rm (B)} (y f^2 / |X|)
  + 2 N_C V_T^{\rm (F)} (y f^2 / |X|)
  - a_L \alpha_3^2 (T) T^4 \ln |X|^2
  \nonumber \\ &&
  + 2 V_T^{\rm (B)} (\lambda \sqrt{|X|^2 + f^4 / |X|^2})
  + 2 V_T^{\rm (F)} (\lambda \sqrt{|X|^2 + f^4 / |X|^2}).
  \label{V_T(X2)}
\end{eqnarray}
Here, $\tilde{m}_i$ are effective masses which may include the effect
of Hubble induced masses; $\tilde{m}_i\simeq m_i$ when $H\ll m_{3/2}$,
while $\tilde{m}_i\sim H$ if the Hubble induced masses exist and
dominate.  In Eq.\ \eqref{V_T(X2)}, $N_c=3$ is the color factor.  In
addition, in our numerical study, we take $a_L=1$.

Once the amplitudes of $\bar{X}$ and $X$ become non-vanishing,
$\hat{Q}_0$ and $\hat{Q}_+$ become massive with the mass of
$y|\bar{X}|$.  In addition, $\hat{S}$ and $\hat{\cal X}$ acquire the
mass of $\lambda \sqrt{|X|^2 + |\bar{X}|^2}$.  After the PQ symmetry
breaking, the product $\bar{X}X$ is fixed to be
$f^2+O(yT\tilde{m}_1)$.  Thus in the situation of our interest, we can
eliminate $\bar{X}$ (or $X$) from the potential.  We choose $X$ as the
independent variable because $\hat{X}$ plays the role of the axion
multiplet when $X\gg \bar{X}$, which is the case in the following
discussion.

Although $\hat{S}$ and $\hat{\cal X}$ also affect the thermal
potential, it is instructive to consider the case that the thermal
effects are dominated by those of $\hat{Q}_0$ and $\hat{Q}_+$.  In
such a case, the potential is approximated as
\begin{eqnarray}
  V_T \sim \tilde{m}_1^2 |X|^2 + \frac{\tilde{m}_2^2 f^4}{|X|^2}
  + a_T y^2 T^2 \frac{f^4}{|X|^2} \theta (T - yf^2/|X|)
  - a_L \alpha_3^2 T^4 \ln |X|^2 ,
\end{eqnarray}
where $a_T$ is a constant of $O(1)$.  When the temperature is high
enough, the potential have a minimum at $X\sim x_T$, where
\begin{eqnarray}
  x_T \equiv \sqrt{\frac{yT}{\tilde{m}_1}} f.
  \label{x_T}
\end{eqnarray}
(Hereafter, we call the minimum at $X\sim x_T$ as ``trapping
minimum.'')  The trapping minimum disappears when $T$ becomes smaller
than the effective mass of the PQ fermions at $X\sim x_T$.  This
happens when the cosmic temperature is $\sim T_{\rm c}$, where $T_{\rm
  c}$ is the solution of the following equation:
\begin{eqnarray}
  T_c = \frac{y f^2}{x_T(T_c)}.
  \label{T_c}
\end{eqnarray}
(Notice that, when the Hubble-induced mass is sizable, $\tilde{m}_1$
depends on background temperature.)  On the contrary, at low
temperature, the $T^2$ term in the potential does not exist at the
region with $X\ll yf^2/T$.  If so, there may exist a minimum at $X\sim
x_L$, where $x_L$ is given by
\begin{eqnarray}
  x_L^2 \equiv 
    \frac
    {a_L \alpha_3^2 T^4 + 
      \sqrt{(a_L \alpha_3^2 T^4)^2 + 4 \tilde{m}_1^2 \tilde{m}_2^2 f^4}}
    {2 \tilde{m}_1^2}.
\end{eqnarray}
When $a_L \alpha_3^2 T^4\gtrsim \tilde{m}_1\tilde{m}_2 f^2$, $x_L\sim
a_L^{1/2}\alpha_3 T^2/\tilde{m}_1$, which can be significantly larger
than $f$.  Behavior of the potential at $X\sim x_L$ depends on the
effective mass of the PQ fermion at such a field value.  If the PQ
fermion mass is smaller than $T$, $\hat{Q}_0$ and $\hat{Q}_+$ (and
their scalar partners) are thermalized as relativistic particles at
$X\sim x_L$, and hence the $T^2$ term in the potential exists.  In
such a case, $X\sim x_L$ is not the minimum of $V_T$, but there exists
only one minimum at $X\sim x_T$.  This is the case when $T\gtrsim
T_*$, where $T_*$ is obtained by solving the following equation:
\begin{eqnarray}
  T_* = \frac{y f^2}{x_L(T_*)}.
  \label{T_*}
\end{eqnarray}
(In defining $T_*$, effects of $\hat{S}$ and $\hat{\cal X}$ are not
included; their effects will be separately considered later).  If the
Hubble-induced mass is negligible, $T_*=\alpha_3^{-1/3} y^{1/3}
m_1^{1/3} f^{2/3}$ (where we have assumed $\alpha_3 y^2
(f/m_{3/2})^5\gtrsim 1$, which is the case in the situation of our
interest.)  On the contrary, for $T\lesssim T_*$, the PQ fermions are
heavier than $T$ at $X\sim x_L$, and we expect that the potential has
a minimum at $X\sim x_L$.

In summary, the thermal potential behaves as follows:
\begin{itemize}
\item $T\lesssim T_{\rm c}$: There is only one minimum at $X\sim x_L$.
\item $T_{\rm c}\lesssim T\lesssim T_*$: There are two minima at
  $X\sim x_L$ and $X\sim x_T$.
\item $T\gtrsim T_*$: There is only one minimum at $X\sim x_T$.
\end{itemize}
We should also comment here that, if $x_L(T_{\rm c})\sim O(f)$, two
minima become indistinguishable at $T\sim T_{\rm c}$.  In such a case,
trapping minimum, which exists at high temperature, smoothly merges to
the zero temperature minimum. Our numerical calculation indicates that
this happens when $y\lesssim 7 \sqrt{m_2/f}$.  In this case, the
coherent oscillation of the saxion is not induced.

\begin{figure}
  \centerline{\epsfxsize=\textwidth\epsfbox{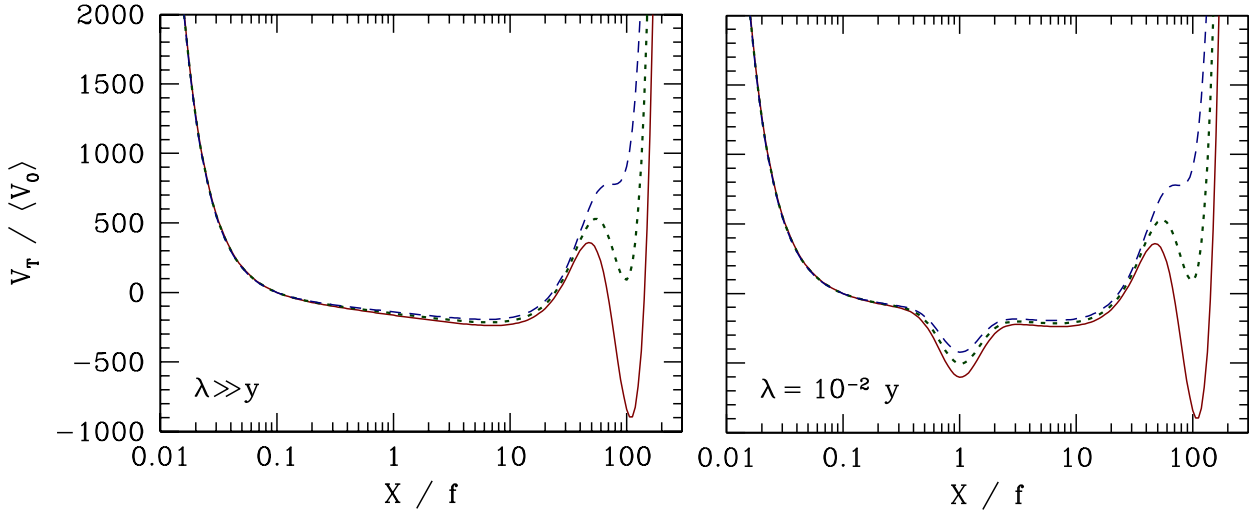}}
  \caption{\small Thermal potential as a function of $X/f$.  We take
    $f=10^{10}\ {\rm GeV}$, $m_1=m_2=1\ {\rm TeV}$, and $y=1$.  The
    solid (dotted, dashed) line is for $T=0.69T_c$, ($T=0.67T_c$,
    $T=0.65T_c$).  We added a constant to the potential.  In addition,
    the vertical axis is normalized by the zero-temperature
    expectation value of the potential $\langle V_0\rangle\equiv
    m_1^2\langle X\rangle^2+m_2^2\langle \bar{X}\rangle^2$.  The left
    figure is for the case that the contributions of $\hat{S}$ and
    $\hat{\cal X}$ are negligible, while the right figure is for the
    case with $\lambda=0.01$.  }
  \label{fig:potshape}
\end{figure}

In Fig.\ \ref{fig:potshape}, we plot $V_T$ given in Eq.\
\eqref{V_T(X2)} for the temperature around $T\sim T_c$, neglecting the
effects of $\hat{S}$ and $\hat{\cal X}$.  (Here, we take $\tilde{m}_i=
m_i$ and the effects of Hubble-induced masses are omitted.)  As one
can see, when the temperature is $O(T_c)$, the trapping minimum
disappears.  In addition, two distinct minima exist just before the
disappearance of the trapping minimum.  Behaviors shown in the figure
are consistent with the previous discussion.

So far, we have considered the case that the thermal effects from
$\hat{S}$ and $\hat{\cal X}$ are negligible.  However, if $\lambda
f\lesssim T_c$, the potential around $X\sim f$ is deformed.  In
particular, if $\lambda$ is sizable, there exists a minimum at $X\sim
f$ at $T\sim T_*$.  In Fig.\ \ref{fig:potshape}, we also show one of
the examples of such a case, taking $\lambda=0.01y$.  Such an extra
minimum may affect the evolution of the saxion, in particular the
trapping process.  (See the following discussion.)

The thermal trapping of the saxion may be cosmologically important,
although its implication strongly depends on the details of the model.
In the following, we discuss possible effects of the thermal trapping
in the present model.  We pay particular attention to the energy
density of axion produced from the saxion oscillation.  If the saxion
field is once trapped in the trapping minimum, it starts to oscillate
with the amplitude of $\sim x_T$ at $T\sim T_{\rm c}$.  Then, such a
saxion oscillation decays with long lifetime.  If $\sigma$ dominantly
decays into axion pair, the produced axion survives until today and
behaves as dark radiation whose abundance is constrained from the
cosmic microwave background (CMB) anisotropy and big-bang
nucleosynthesis (BBN).

\subsection{Case with $y<\lambda$}

Thermal history depends on the mass spectrum of particles in the PQ
sector.  We first consider the case with $y<\lambda$; in such a case,
$\hat{S}$ and $\hat{\cal X}$ are always heavier than $\hat{Q}_0$ and
$\hat{Q}_+$.  Then, at $T\lesssim\lambda f$, we neglect the thermal
effects from $\hat{S}$ and $\hat{\cal X}$.  The cosmological
implication of the saxion depends also on its interactions.  Thus, for
simplicity, we mostly concentrate on the case that the interaction of
the saxion with the Higgses given in Eq.\ \eqref{XXHH} is negligible.

We start our discussion with studying the behavior of $T_*$, which
depends on the size of the Hubble-induced mass.  In the following
study, we consider the case that the Hubble-induced mass is sizable
and approximate
\begin{eqnarray}
  \tilde{m}_i=\mbox{max} (m_i,H).
\end{eqnarray}
The Hubble-induced mass relaxes the constraint compared to the case
with $\tilde{m}_i=m_i$.

\begin{figure}
  \centerline{\epsfxsize=\textwidth\epsfbox{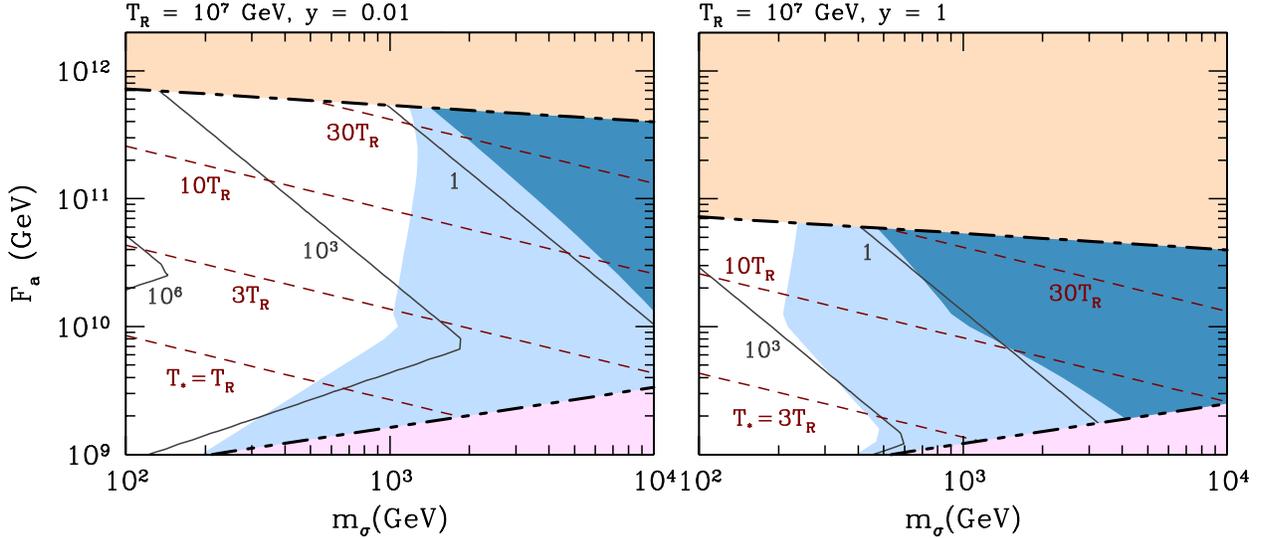}}
  \caption{\small The dotted lines are contours of constant $T_*$.
    The solid lines are contours of constant $R_{\sigma /r}$ (see Eq.\
    \eqref{R(sig/r)}) for $\xi\sim 1$.  (Numbers in the figures are
    the values of $R_{\sigma /r}$.)  In the darkly-shaded (dark blue)
    region, $\Delta N_\nu$ becomes smaller than $1$ if
    $\Gamma_{\tilde{a}}/\Gamma_{\sigma}=10^{-6}$.  In the
    lightly-shaded (light blue) region, $\Delta N_\nu$ becomes smaller
    than $1$ if the axino decays just before the BBN.  Above the
    dot-dashed line, Eq.\ \eqref{T_*} does not have solution.  Below
    the dot-dot-dashed line, $\Gamma_{\rm diss}>H$ is realized at a
    certain epoch after $T=T_{\rm c}$.  Here, $T_{\rm R}=10^7\ {\rm
      GeV}$, and the Yukawa coupling constant is taken to be $y=0.01$
    (left) and $1$ (right).}
  \label{fig:mvsfTR7}
\end{figure}

\begin{figure}
  \centerline{\epsfxsize=\textwidth\epsfbox{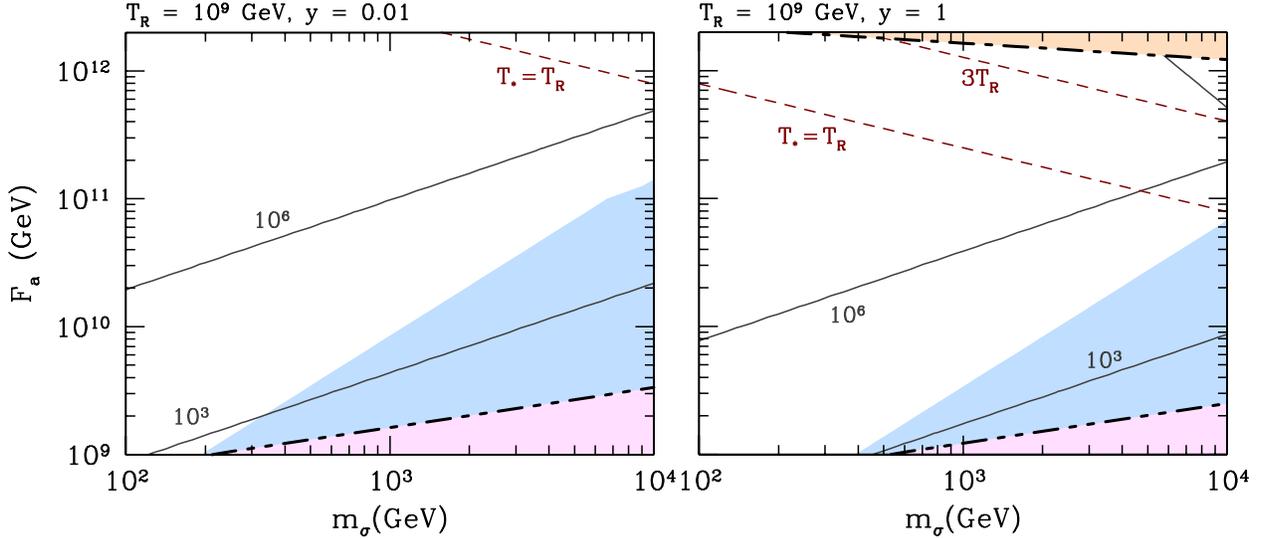}}
  \caption{\small Same as Fig.\ \ref{fig:mvsfTR7}, except for $T_{\rm
      R}=10^9\ {\rm GeV}$.  (No dark shaded region on this parameter
    region.)}
  \label{fig:mvsfTR9}
\end{figure}

In Figs.\ \ref{fig:mvsfTR7} and \ref{fig:mvsfTR9}, we plot the
contours of constant $T_*$ on $m_\sigma$ vs.\ $F_a$ plane for fixed
values of $y$ and $T_{\rm R}$.  (Here and hereafter, we approximate
$f\sim F_a$ and $m_1\sim m_2\sim m_\sigma$.)  The behavior of $T_*$
can be understood as follows.  For small enough $F_a$, effect of the
Hubble-induced mass is irrelevant in solving Eq.\ \eqref{T_*}, and
$T_*$ is independent of $T_{\rm R}$.  On the contrary, for larger
value of $F_a$, effect of the Hubble-induced mass becomes important.
Let us consider how $T_*$ behaves when $\tilde{m}_i\sim H$.  In the
radiation-dominated universe, $H$ is proportional to $T^2$, and hence
$x_L\simeq a_L^{1/2}\alpha_3 T^2/H$ becomes insensitive to $T$.  In
such a case, $T_*$, which is the solution of Eq.\ \eqref{T_*}, becomes
independent of $m_1$.  On the contrary, in the inflaton dominated
universe, $H\propto a^{-3/2}$ while $T\propto a^{-3/8}$, and hence
\begin{eqnarray}
  H(T>T_{\rm R}) \simeq
  \left( \frac{T}{T_{\rm R}} \right)^4
 H(T_{\rm R}),
\end{eqnarray}
where $T$ is the temperature of dilute plasma produced by the decay of
inflaton.  Then, $x_L^{-1}$ is proportional to $T^2$ as the
temperature of dilute plasma increases, and Eq.\ \eqref{T_*} does not
have a solution if $F_a$ is too large.  Consequently, the thermal
trapping of the saxion is not guaranteed if $F_a\gtrsim
g_*^{-3/16}y^{-1/2}\alpha_3^{1/2}m_\sigma^{1/8}T_{\rm R}^{3/4}M_{\rm
  Pl}^{3/8}$.  We conservatively assume that no constraint is obtained
in such a case.

If the saxion is once trapped in the trapping minimum, the saxion
starts to oscillate with the amplitude of $O(x_T)$ when the trapping
minimum disappears.  Even if $X$ and $\bar{X}$ are of $O(f)$ just
after the PQ phase transition, we expect that the thermal trapping
occurs if $T_*$ is lower than the maximum temperature of the universe
$T_{\rm max} \sim H_{\rm inf}^{1/4}M_{\rm Pl}^{1/4}T_{\rm R}^{1/2}$,
with $H_{\rm inf}$ being the expansion rate during inflation
\cite{Kolb:1990vq}.

In Figs.\ \ref{fig:mvsfTR7} and \ref{fig:mvsfTR9}, we also show the
region in which the dissipation rate becomes larger than $H$ at a
certain cosmic temperature below $T_{\rm c}$, where the dissipation
rate is evaluated with Eq.\ \eqref{Gamma_diss} by replacing $F_a$ with
the amplitude of $X$.  In such a region, even if the trapping happens,
the saxion oscillation dissipates away.  Then, the relic saxion is
dominated by thermally produced one.  Notice that the region with
$y\lesssim 7 \sqrt{m_2/f}$, in which two minima merge smoothly, is
mostly covered by the region with $\Gamma_{\rm diss}>H$.  One may
think that the scattering processes among the particles in the PQ
sector also contribute to the dissipation process.  At $T<T_c$, only
the particles in the axion multiplet $\hat{\cal A}$ (i.e., axion,
axino, and saxion) can contribute to the dissipation process because
the masses of other particles in the PQ sector are larger than $T_c$.
The axion multiplet particles are thermalized only when $\Gamma_{\rm
  diss}$ becomes comparable to $H$.  In addition, the interaction of
the axion multiplet is suppressed by inverse powers of $F_a^{\rm
  (eff)}$ because the axion is a Nambu-Goldstone boson.  Dissipation
rate due to the scattering processes among the PQ sector particles is
estimated to be $\sim \frac{1}{4\pi}\frac{T^5}{F_a^{{\rm (eff)}4}}$
and is smaller than $\Gamma_{\rm diss}$ given in Eq.\
\eqref{Gamma_diss} (with $F_a\rightarrow F_a^{\rm (eff)}$) in the
parameter region of our interest.

If the dissipation of the saxion oscillation is negligible, the saxion
oscillation with large initial amplitude may result in the
overproduction of the axion dark radiation.  To see when there may
exist such a problem, we calculate the density of the oscillating
saxion.  The yield variable of the saxion is estimated as
\begin{eqnarray}
  Y_\sigma^{\rm (osc)} \simeq 
  \frac{45}{\pi^2}
  \frac{\tilde{m}_\sigma \sigma_{\rm R}^2}{g_*(T_{\rm R}) T_{\rm R}^3},
\end{eqnarray}
where $\sigma_{\rm R}$ is the saxion amplitude at $T=T_{\rm R}$, and
$\tilde{m}_\sigma$ is the effective mass of saxion which may include
the effect of the Hubble-induced mass.  In our numerical analysis, we
use $g_*(T_{\rm R})=228.75$.  (If the saxion starts to oscillate after
the reheating, $T_{\rm R}$ should be replaced by the temperature at
which the oscillation starts, and $\sigma_{\rm R}$ should be
identified as the initial amplitude of the saxion oscillation.)
If the oscillation of the saxion starts before the reheating,
$n_\sigma(T_{\rm R})$ is related to the initial number density using
the fact that $a\propto T^{-3/8}$ during the inflaton-dominated era.
Denoting the temperature at which the saxion starts to oscillate as
$T_i$ (with $T_i>T_{\rm R}$),
\begin{eqnarray}
  n_\sigma (T_{\rm R}) \simeq \left( \frac{T_{\rm R}}{T_i} \right)^8
  n_\sigma (T_i).
\end{eqnarray}

To see how large the saxion abundance is, we calculate the following
ratio
\begin{eqnarray}
  R_{\sigma /r} \equiv 
  \frac{\rho_\sigma(t_{\rm dec})}{\rho_r(t_{\rm dec})},
  \label{R(sig/r)}
\end{eqnarray}
where $\rho_\sigma(t_{\rm dec})$ and $\rho_r(t_{\rm dec})$ are energy
densities of the saxion and the radiation from the inflaton decay at
the time just before the saxion decay, respectively.  If $R_{\sigma
  /r}\gtrsim 1$, the saxion dominates the universe.  In Figs.\
\ref{fig:mvsfTR7} and \ref{fig:mvsfTR9}, we show contours of constant
$R_{\sigma /r}$.  We can see that the saxion dominance occurs in large
fraction of the parameter space.  Then, when it decays, some amount of
energetic axion is produced.

The present density of the axion from the saxion decay is parametrized
by using the effective number of extra neutrinos as
\begin{eqnarray}
  \Delta N_\nu \equiv N_\nu^{\rm (eff)} - 3 \equiv 
  \frac{3\rho_{a}(t_{\rm now})}{\rho_{\nu}(t_{\rm now})}.
\end{eqnarray}
Here, $\rho_{a}(t_{\rm now})$ and $\rho_{\nu}(t_{\rm now})$ are energy
densities of the axion and neutrinos in the present universe,
respectively, and the factor of $3$ in the numerator is the number of
generations of neutrinos.  From the latest WMAP observation
\cite{Komatsu:2010fb}, the best-fit value of the effective number of
neutrinos (which includes the standard-model contribution) is obtained
as $N_\nu^{\rm (eff)}=4.34^{+0.86}_{-0.88}$ (68\% C.L.).  In addition,
the analysis of $^4{\rm He}$ mass fraction generated by the BBN
reactions indicates $N_\nu^{\rm (eff)}=3.68^{+0.80}_{-0.70}$
($2\sigma$) \cite{Izotov:2010ca}.  

If the saxion abundance is too large, $\Delta N_\nu$ exceeds $\sim 1$,
which conflicts with observations.  In estimating $\Delta N_\nu$, we
should notice that the entropy production due to the axino decay may
be sizable in the present case.  $\Delta N_\nu$ becomes smaller as the
decay rate of axino $\Gamma_{\tilde{a}}$ gets smaller.  The axino
decays into gaugino and gauge boson pair and the total decay rate of
$\tilde{a}$ is smaller than that of $\sigma$ if $\xi$ is close to $1$.
Typically, $\Gamma_{\tilde{a}}/\Gamma_{\sigma}\sim O(N_i
\alpha_i^2/2\pi^2\xi^2)$, where $N_i$ is the number of final states
and $i$ depends on the dominant decay mode; if the axino mass is well
above the gluino mass, $N_i=8$ and $\alpha_i=\alpha_3$.  However, if
the masses of the axino and the gauginos are degenerate,
$\Gamma_{\tilde{a}}$ becomes suppressed.  So the ratio
$\Gamma_{\tilde{a}}/\Gamma_{\sigma}$ is model dependent.  Notice that,
if $\Gamma_{\tilde{a}}$ is too small, the axino decays after the BBN,
which is likely to spoil the success of the standard BBN scenario.  We
vary $\Gamma_{\tilde{a}}$ and estimate $\Delta N_\nu$.  (We use the
instantaneous decay approximation.)  Here, the axino abundance is
evaluated using Eq.\ \eqref{Y(axino)}, and the axino mass is taken to
be $\frac{1}{2}m_\sigma$.  In Figs.\ \ref{fig:mvsfTR7} and
\ref{fig:mvsfTR9}, we show the region in which $\Delta N_\nu$ becomes
smaller than $1$ if $\Gamma_{\tilde{a}}/\Gamma_{\sigma}=10^{-6}$, and
also the region in which $\Delta N_\nu$ becomes smaller than $1$ if
the axino decays just before the BBN. If
$\Gamma_{\tilde{a}}/\Gamma_{\sigma}=10^{-6}$, $\Delta N_\nu\gtrsim 1$
for $R_{\sigma/r}\gtrsim 1$ even with the entropy production due to
the axino decay.  If the axino decays just before the BBN, the effect
of the entropy production is more significant.  Here, the energy
density stored in the axino sector is assumed to be fully converted to
that of radiation after the axino decay.  This may be due to the pair
annihilation of the LSP or due to the decay of the LSP via $R$-parity
violation.  Even with such an extreme assumption, $\Delta N_\nu$ may
be larger than $\sim 1$ if $Br(\sigma\rightarrow aa)\simeq 1$ in
significant fraction of the parameter space.

Such a problem may be avoided if the SUSY breaking parameters are
tuned so that the $\xi$ parameter becomes relatively small.  Indeed,
if $\xi\lesssim 0.05$, $Br(\sigma\rightarrow gg)$ becomes comparable
to or larger than $Br(\sigma\rightarrow aa)$, and the production of
the axion is suppressed.  In the region with $R_{\sigma /r}\gtrsim 1$
of the figure, saxion once dominates the universe. Even so, $\Delta
N_\nu\lesssim 1$ is possible if $Br(\sigma\rightarrow aa) \sim
O(0.1)$.  In fact, as one can see, the constraints on $N_\nu^{\rm
  (eff)}$ from the CMB anisotropy and BBN indicate non-vanishing value
of $\Delta N_\nu$; $\Delta N_\nu\sim 1$ is preferred.  If
$Br(\sigma\rightarrow aa) \simeq 0.25$, $\Delta N_\nu =1$ is realized
(with $g_*=100$ at the time of saxion decay).  Another solution is to
introduce the superpotential interaction given in Eq.\ \eqref{XXHH}.
With such an interaction, $\sigma$ may dominantly decay into Higgs
boson pair, and $Br(\sigma\rightarrow aa)$ can be $O(0.1)$.\footnote
{The interaction with the Higgses may enhance the dissipation rate,
  which may also help to reduce the density of the axion.  The study
  of the dissipation rate in such a case will be given elsewhere
  \cite{MorTak}.}
In such a case, the saxion may also decay into the Higgsino pair,
which may result in the overproduction of relic LSP.  However, such a
difficulty can be avoided with large enough pair annihilation cross
section of the LSP.  This is the case, for example, if the LSP is the
neutral Wino \cite{Moroi:1999zb}.  If the saxion dominantly decays
into MSSM particles, one should take account of the entropy production
due to the saxion decay.  The dilution factor is $\sim
R_{\sigma/r}^{3/4}$.

\subsection{Case with $y>\lambda$}

Next, we consider the case with $y>\lambda$.  In such a case, if
$X\sim \bar{X}\sim f$, $\hat{Q}_0$ and $\hat{Q}_+$ become heavier than
$\hat{S}$ and $\hat{\cal X}$.  Even so, the trapping may happen if
$T_*\lesssim \lambda f$; in such a case, discussion in the previous
section holds.

If $T_*\gtrsim \lambda f$, the situation is rather complicated.  This
is because $\hat{S}$ and $\hat{\cal X}$ may be fully thermalized as
relativistic particles at $T\lesssim T_c$.  Then, the scattering
processes with $\hat{S}$ and $\hat{\cal X}$ significantly contributes
to the dissipation of the oscillation of the saxion.  In addition, if
$T_*\gtrsim \lambda f$, there may exist a minimum at $X\sim f$ caused
by the thermal effects of $\hat{S}$ and $\hat{\cal X}$, with which the
trapping process to the trapping minimum may not happen.  (See Fig.\
\ref{fig:potshape}.)  The $\hat{S}$ and $\hat{\cal X}$ can be produced
from the scattering of particles in thermal bath (like gluon).  Even
though the interaction between the MSSM sector and the PQ sector
(which consists of $\hat{X}$, $\hat{\bar{X}}$ and $\hat{S}$) may be
too weak to equate the temperatures of two sectors, we expect that the
interactions among $\hat{X}$, $\hat{\bar{X}}$ and $\hat{S}$ are strong
enough to thermalize the PQ sector; such a thermalization occurs
through the superpotential interaction of
$\lambda\hat{S}\hat{\bar{X}}\hat{X}$.  Thus, if the masses of
$\hat{S}$ and $\hat{\cal X}$ are smaller than $T_*$ (at $X\sim f$),
constraint obtained in the previous section may not be applicable.
This is the case when $\lambda\lesssim y^{1/3}(m_\sigma/F_a)^{1/3}$.

\subsection{Saxion from Scattering of Thermal Particles}

Finally, we comment on the effects of thermally produced saxion.  As
we have mentioned, scattering processes of thermal particles produce
saxion.  In particular, if $\Gamma_{\rm diss}\sim H$ is realized at
some epoch, the saxion are thermalized.  In such a case, the
overproduction of the axion may occur if the saxion dominantly decays
into axion pair.  However, the abundance of thermally produced axino
is comparable to that of saxion.  Then, entropy production due to the
decay of axino (as well as due to the subsequent pair annihilation of
the LSP) dilutes the axion abundance.  If the decay rate of the axino
is smaller than that of saxion, the effect of the dilution is large
enough to make $\Delta N_\nu$ smaller than $\sim 1$.

\section{Discussion}
\label{sec:discussion}
\setcounter{equation}{0}

In this paper, we have studied effects of saxion on the evolution of
the universe paying particular attention to thermal effects on the
saxion.  Because the axion multiplet necessarily couples to colored
particles (i.e., PQ fermions), saxion potential is inevitably deformed
at high temperature.  In addition, the coherent oscillation of the
saxion may dissipate via the interaction with hot plasma.

Taking account of these effects, we have studied the constraint on the
supersymmetric PQ model with the superpotential given in Eq.\
\eqref{superpot}.  Because of the thermal effect on the potential, the
saxion field may be trapped in a false minimum before the start of the
oscillation, which may result in an overproduction of relativistic
axion.  The constraint strongly depends on the model parameters.  We
have seen that, if no PQ sector particle remains in thermal bath after
the onset of the oscillation, $\Delta N_\nu\gg 1$ may happen if
$Br(\sigma\rightarrow aa)\simeq 1$.  In addition, if the dissipation
rate is large enough, the saxion oscillation dissipates soon after the
start of the oscillation.

In this study, we have concentrated on the model with the
superpotential given in Eq.\ \eqref{superpot}.  However, the thermal
effects may have significant effects in other classes of models.

\noindent {\it Acknowledgements}: The authors thank K. Hamaguchi,
K. Mukaida and M. Nagai for useful discussion.  The work of T.M. is
supported by Grant-in-Aid for Scientific research from the Ministry of
Education, Science, Sports, and Culture (MEXT), Japan, No.\ 22244021,
No.\ 23104008, and No.\ 60322997.

\end{document}